\documentclass[%
 reprint,
nofootinbib,
 amsmath,amssymb,
 aps,
prl,
]{revtex4-2}

\usepackage{graphicx}
\usepackage{dcolumn}
\usepackage{bm}
\usepackage{hyperref}

\begin{document}


\title{Wave-Optics Imprints of Dark Matter Subhalos \\ on Strongly Lensed Gravitational Waves}

\author{Shin'ichiro Ando}
\affiliation{GRAPPA Institute, University of Amsterdam, 1098 XH Amsterdam, The Netherlands}
\affiliation{Kavli Institute for the Physics and Mathematics of the Universe (WPI), The University of Tokyo, Kashiwa 277-8583, Japan}

\date{March 4, 2026}

\begin{abstract}
Wave-optics effects in strongly lensed gravitational waves (GWs) provide a new interferometric probe of dark matter substructure. 
We compute the full diffraction integral for GWs propagating through statistically generated cold dark matter subhalo populations and quantify the resulting frequency-dependent amplification in the Laser Interferometer Space Antenna (LISA) band. 
We show that realistic galaxy-scale lenses generically produce percent-level amplitude and phase distortions in strongly magnified images, primarily induced by subhalos in the mass range $10^4$--$10^7\,M_{\odot}$. 
These signatures arise naturally within the standard cold dark matter paradigm and should be detectable in high signal-to-noise LISA events. 
Strongly lensed GWs thus offer a direct and complementary window on dark matter structure at subgalactic mass scales inaccessible to electromagnetic measurements.
\end{abstract}

\maketitle


\section{\label{sec:Introduction}Introduction}


Dark matter substructure is a fundamental prediction of hierarchical structure formation and a sensitive probe of the particle nature of dark matter~\cite{Zavala:2019gpq}. In the standard cold dark matter paradigm, galactic halos are expected to host a large population of bound subhalos spanning many orders of magnitude in mass~\cite{Springel:2008cc, Hiroshima:2018kfv}. Testing this prediction at small scales is crucial for distinguishing between cold, warm, self-interacting, or wave-like dark matter scenarios~\cite{Bullock:2017xww, Lovell:2013ola,Dekker:2021scf,Tulin:2017ara,Ando:2024kpk,Hui:2016ltb}. However, current observational probes such as stellar streams~\cite{Banik:2019cza}, strong lensing flux anomalies~\cite{Vegetti:2023mgp}, and dwarf galaxy counts and velocity dispersions~\cite{Simon:2019nxf,Nadler:2021dft,Ando:2025qtz} remain limited in sensitivity and are subject to significant astrophysical uncertainties. Identifying robust and complementary methods to detect dark matter subhalos therefore remains an outstanding challenge.


Gravitational lensing of gravitational waves (GWs) provides a qualitatively new probe of dark matter structure, as phase coherence leads to frequency-dependent interference effects when the characteristic lensing time delay is comparable to the GW period~\cite{Takahashi:2003ix}. Such wave-optics (WO) effects have been extensively studied over the past two decades, particularly in the context of compact objects and low-mass perturbers in both ground- and space-based detectors~\cite{Jung:2017flg,Lai:2018rto,Tambalo:2022wlm,Caliskan:2023zqm,Villarrubia-Rojo:2024xcj}. Forecasts for the Laser Interferometer Space Antenna (LISA) and Laser Interferometer Gravitational-Wave Observatory (LIGO) bands that assume generic line-of-sight halo configurations~\cite{Fairbairn:2022xln,Guo:2022dre,Brando:2024inp} suggest that detectable frequency-dependent modulations are expected to occur only in a limited number of events over the mission lifetime. Similar conclusions extend to next-generation detectors such as the Einstein Telescope, reinforcing the view that WO signatures might be observationally challenging in unselected lensing geometries.


However, strongly lensed GW systems represent a qualitatively different regime. By construction, such events probe lines of sight that pass through the inner regions of massive foreground halos, where the projected abundance of dark matter subhalos is high. In addition, near the critical curves of the macrolens, small-scale perturbations are geometrically amplified, enhancing the response of the lensed GW waveform to subhalo-induced distortions relative to generic lines of sight. The large macroscopic magnification further increases the signal-to-noise ratio, improving the detectability of frequency-dependent modulations. Finally, strong lensing itself is expected to occur at a non-negligible rate in the LISA band~\cite{Sereno:2010dr,Oguri:2018muv,Gutierrez:2025ymd}, implying that multiply imaged GW events provide a well-defined and physically motivated sample in which WO signatures of dark matter substructure can be systematically searched.


In this \textit{Letter}, we demonstrate that strongly lensed GW events provide a realistic and generically favorable setting for observing WO imprints from dark matter subhalos. By modeling GW propagation through a macroscopic strong-lensing potential populated with a physically motivated distribution of subhalos, we compute the full frequency-dependent amplification factor in the WO regime. We show that percent-level modulations in the mHz GW waveform arise naturally for subhalo masses in the range $\sim 10^{4}$--$10^{7}\,M_\odot$ when the source lies near a macro-critical curve. Crucially, these signatures occur without requiring extreme impact parameters or isolated compact objects. For typical LISA massive black-hole binaries with signal-to-noise ratios $\mathrm{S/N} \gtrsim 100$, the predicted distortions are detectable, implying that strongly lensed GW events constitute a promising new probe of dark matter substructure.


\section{Model and Formalism}

We consider a strongly lensed GW system consisting of a macroscopic host halo populated with dark matter subhalos. Subhalo populations are generated using the semi-analytic SASHIMI model~\cite{Hiroshima:2018kfv,Ando:2019xlm,Ando:2020yyk},\footnote{\url{https://github.com/shinichiroando/sashimi-c}}
which self-consistently predicts the subhalo mass function and structural evolution within a host halo.
Each subhalo is modeled with a truncated Navarro-Frenk-White (NFW) density profile characterized by characteristic density $\rho_s$, scale radius $r_s$, and truncation radius $r_t$ beyond which the density profile is strongly suppressed due to tidal stripping.
All WO lensing calculations are performed with the GLoW framework~\cite{Villarrubia-Rojo:2024xcj},\footnote{\url{https://github.com/miguelzuma/GLoW_public}}
which provides a flexible implementation of lensing potentials and numerical evaluation of both time- and frequency-domain WO amplification factors. 

\subsection{Macrolens model}

The macrolens consists of a dark matter halo at redshift $z_L = 0.5$ described by an NFW density profile with $M_{200\mathrm{c}} = 10^{12}\,M_\odot$ (mass defined within a radius $R_{200\mathrm{c}}$ enclosing a mean density equal to 200 times the critical density at $z_L$) and concentration $c_{200\mathrm{c}} = 10/(1+z_L)$~\cite{Correa:2015dva}. A central galaxy is modeled as a singular isothermal sphere (SIS) with velocity dispersion $\sigma_v = 250\,\mathrm{km\,s^{-1}}$.
Subhalos with masses $m>10^9\,M_\odot$ are included within the macrolens model following a spatial distribution proportional to $(r^2 + R_s^2)^{-3/2}$, where $R_s = R_{200\mathrm{c}}/c_{200\mathrm{c}}$ is the NFW scale radius of the host. 
An example realization of the massive subhalo population is shown in the inset of Fig.~\ref{fig:subs_dist} (see Appendix~A for detailed procedure).
We consider a GW source at redshift $z_S = 1.5$, located at a dimensionless source position $y_{\rm src}=0.1$, where lens-plane coordinates are normalized by the Einstein radius of the SIS component.

\begin{figure}[t]
\centering
\includegraphics[width=\columnwidth]{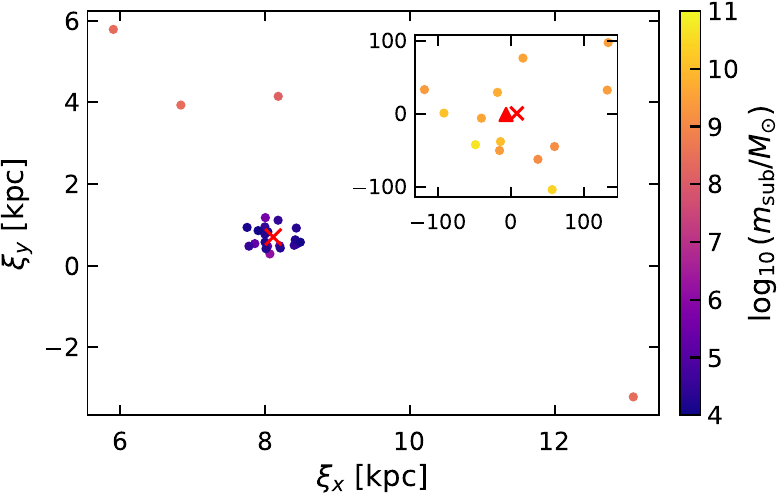}
\caption{
Spatial distribution of dark-matter subhalos in a representative realization of the strongly lensed system.
The red cross (triangle) marks the macro minimum (saddle) image.
Colored points in the main panel denote 26 low-mass subhalos ($10^4$--$10^9\,M_\odot$) included explicitly in the WO calculation,
while the inset shows 13 massive subhalos ($m_{\rm sub}>10^9\,M_\odot$) treated in the GO limit.
For visual clarity, subhalos in the mass range $10^2$--$10^4\,M_\odot$ (1218 objects in this realization) are not displayed, although they are included in the WO calculation.
The color scale indicates $\log_{10}(m_{\rm sub}/M_\odot)$.
}
\label{fig:subs_dist}
\end{figure}

The resulting lensing potential $\psi_{\rm macro}$ defines the Fermat surface whose stationary points determine the image configuration. In practice, subhalos act as perturbations to the dominant macrolens potential, and for the vast majority of realizations the system retains a two-image configuration analogous to the smooth case, corresponding to a minimum and a saddle point.
In the LISA band, macro-image time delays far exceed the GW period, so their interference is effectively in the geometric-optics (GO) limit and does not produce observable frequency-dependent modulations. We therefore focus on the minimum image, where any detectable modulation must arise from WO effects induced by subhalos.

\subsection{Subhalos around the minimum image}

We populate the vicinity of the macro minimum image with subhalos in the mass range $10^{2}$--$10^{9}\,M_\odot$. For each subhalo mass $m_{\rm sub}$, we restrict the sampling to a cylindrical region of projected radius $R_{\rm near}$ centered on the minimum image. The selection radius is defined as
\begin{equation}
R_{\rm near}
=
\max\!\left[
N_F R_F,\;
R_{\mu},\;
R_{\rm core}
\right],
\end{equation}
where the three terms correspond to physically distinct scales.

The first term is a multiple ($N_F=5$) of the Fresnel scale $R_F=[{c\, d_{\rm eff}}/{(2\pi f_{\min})}]^{1/2}$, where $d_{\rm eff} = D_LD_{LS}/[(1+z_L)D_S]$ with $D_L$, $D_S$, and $D_{LS}$ denoting the angular-diameter distances to the lens, to the source, and between the lens and source, respectively.
This characterizes the transverse coherence scale of WO effects at the minimum GW frequency $f_{\min} = 10^{-4}\,{\rm Hz}$.
The second term, $R_{\mu}=
[{g_{\rm img}\, m_{\rm sub}}/{(\pi \epsilon_\mu \Sigma_{\rm cr})}]^{1/2}$,
is a perturbation radius obtained by requiring that the subhalo induce at least a fractional magnification perturbation $\epsilon_\mu=10^{-2}$ in the GO limit. Here $\Sigma_{\rm cr}=c^2/[4\pi G(1+z_L)d_{\rm eff}]$ is the critical surface density and
$g_{\rm img}\equiv \|A_{\rm min}^{-1}\|$ encodes the local macro-amplification through the inverse Jacobian of the macrolens evaluated at the minimum image. (See Appendix~B for derivation.)
The third term,
$R_{\rm core}=\min\!\left[N_E r_{s,{\rm sub}},\, r_{t,{\rm sub}}\right]$,
with $N_E=5$, ensures that the sampling radius does not fall below the effective internal scale of the truncated NFW subhalo.

Subhalos are then drawn within this mass-dependent cylindrical region according to the host spatial distribution (Appendix~A). This procedure guarantees inclusion of all perturbers capable of producing detectable WO signatures while avoiding unnecessary sampling of dynamically irrelevant distant subhalos.
A representative realization of the resulting subhalo distribution around the macro minimum image is shown in Fig.~\ref{fig:subs_dist}.

\subsection{WO calculation}

We adopt the standard dimensionless formulation of WO gravitational lensing~\cite{Takahashi:2003ix, Villarrubia-Rojo:2024xcj}.
Introducing physical transverse coordinates $\bm{\xi}$ and $\bm{\eta}$ in the lens and source planes, respectively, we define dimensionless variables
$\bm{x}\equiv\bm{\xi}/\xi_0$ and
$\bm{y}\equiv(\bm{\eta}/\xi_0)(D_L/D_S)$,
where $\xi_0$ is an arbitrary lens-plane scale.
The amplification factor then takes the form
\begin{equation}
F(f)=\frac{w}{2\pi i}
\int d^2x\,
\exp\!\left[i w\, \phi(\bm{x},\bm{y})\right],
\end{equation}
where
$w \equiv f/f_0 \equiv 2\pi f \xi_0^2 /(c d_{\rm eff})$
is the dimensionless frequency and
\begin{equation}
\phi(\bm{x},\bm{y})
=
\frac{1}{2}|\bm{x}-\bm{y}|^2
-
\psi(\bm{x})
\end{equation}
is the dimensionless Fermat potential.

We evaluate the integral in coordinates centered on the macro minimum image located at $\bm{x}_{\rm min}$.
Defining $\bm{u}\equiv\bm{x}-\bm{x}_{\rm min}$ and expanding the macrolens contribution to quadratic order, the integral becomes
\begin{equation}
F(f)=\frac{w}{2\pi i}
\int d^2u\,
\exp\!\left[
i w\left(
\frac{1}{2}\bm{u}^T A_{\rm min} \bm{u}
-
\delta\psi(\bm{u})
\right)
\right],
\label{eq:Ff}
\end{equation}
where $A_{\rm min} = I-\nabla\nabla\psi_{\rm macro}\big|_{\bm{x}_{\rm min}}$ is the Jacobian matrix evaluated at the minimum image and $\delta\psi$ denotes the potential from nearby low-mass subhalos treated explicitly in the WO calculation.
The details of the external-field decomposition are summarized in Appendix~C.

In our numerical implementation, we adopt a convenient normalization corresponding to a reference frequency $f_0=1\,\mathrm{Hz}$, which fixes $\xi_0$.
This choice follows the convention used in the GLoW framework~\cite{Villarrubia-Rojo:2024xcj} and does not affect the physical frequency dependence of the amplification factor.

In the GO limit, each image contributes a frequency-independent magnification; therefore any measurable frequency dependence must originate from WO distortions induced by the local subhalo population.

\begin{figure}[t]
\centering
\includegraphics[width=\columnwidth]{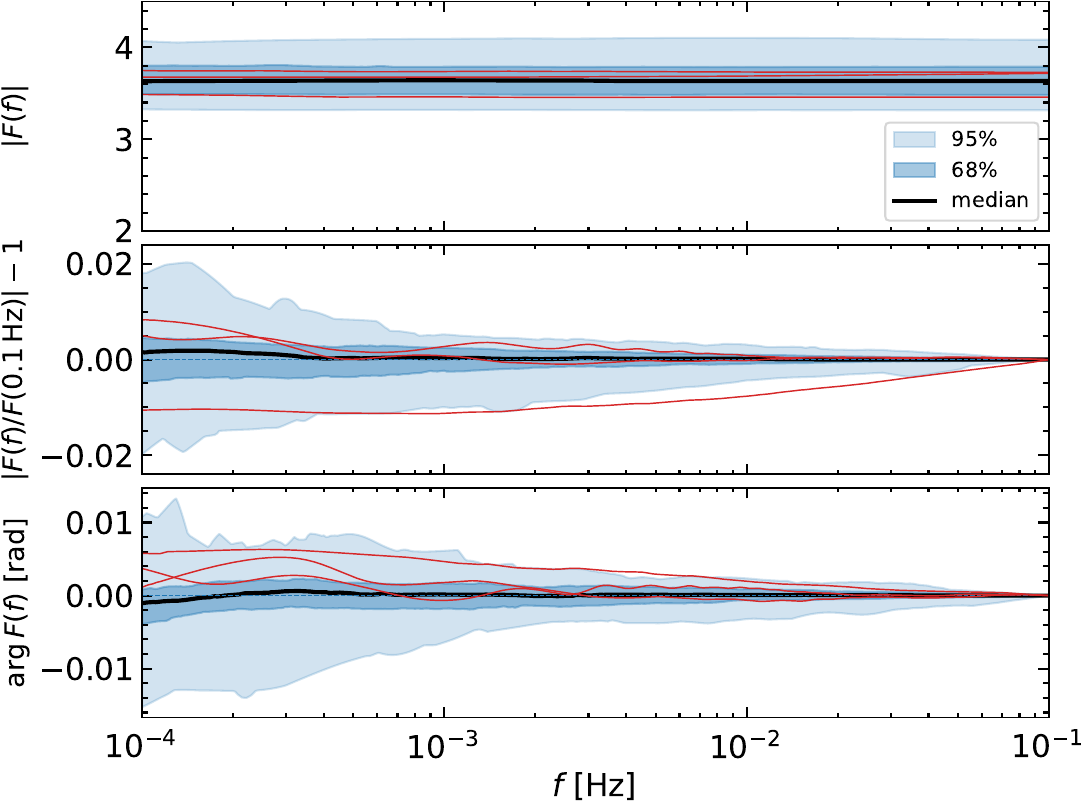}
\caption{
Frequency-dependent gravitational-wave amplification factor in the fiducial strongly lensed configuration.
Top: absolute amplification $|F(f)|$.
Middle: relative amplitude modulation $|F(f)/F(0.1\,\mathrm{Hz})|-1$.
Bottom: phase shift $\arg F(f)$.
Shaded bands indicate the 68\% and 95\% ranges over 200 independent subhalo realizations, and the black curve shows the median.
Thin red curves illustrate three representative realizations.
}
\label{fig:Fw_percentiles}
\end{figure}

\section{WO Signatures}

We now examine the statistical properties of the WO amplification factor in our fiducial strongly lensed configuration.
Figure~\ref{fig:Fw_percentiles} summarizes the frequency-dependent amplification factor over 200 independent subhalo realizations. The top panel shows the absolute amplification $|F(f)|$, the middle panel the relative amplitude modulation $|F(f)/F(0.1,\mathrm{Hz})|-1$, and the bottom panel the phase shift $\arg F(f)$.

The median amplification closely follows the GO expectation, indicating that the dominant macrolens contribution is well described by frequency-independent magnification. However, a non-negligible dispersion is present across realizations. The 68\% and 95\% ranges demonstrate that percent-level amplitude modulations arise in a substantial fraction of realizations, particularly at frequencies $f \lesssim 10^{-3}\, \mathrm{Hz}$. These amplitude variations are accompanied by phase shifts at the level of $\sim 10^{-2}\,\mathrm{rad}$.

\begin{figure}[t]
\centering
\includegraphics[width=\columnwidth]{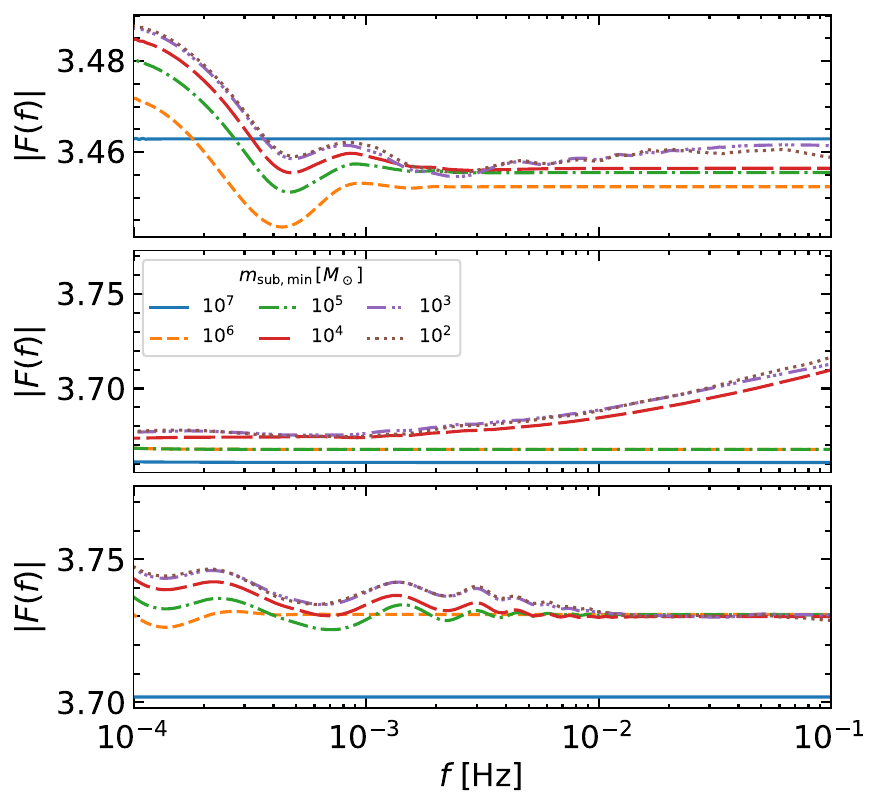}
\caption{
Dependence of the GW amplification factor $|F(f)|$ on the minimum subhalo mass included in the WO calculation.
From top to bottom panels, three realizations (the same as in Fig.~\ref{fig:Fw_percentiles}) are shown.
Different line styles correspond to different minimum subhalo masses 
$m_{\rm sub,min}=10^{2}$--$10^{7}\,M_\odot$.
}
\label{fig:Fw_msubmin}
\end{figure}

To investigate which subhalo mass scales dominate the signal, we vary the minimum subhalo mass included in the WO calculation ($m_{\rm sub}>m_{\rm sub,min}$), as shown in Fig.~\ref{fig:Fw_msubmin}. For each of the three representative realizations highlighted in Fig.~\ref{fig:Fw_percentiles}, the amplification factor is recomputed while progressively lowering the minimum subhalo mass, $m_{\rm sub,min}$. 
We find that the modulation amplitude increases as $m_{\rm sub,min}$ is lowered from $10^{7}$ to $\sim10^{4}\,M_\odot$, but saturates once subhalos down to $\sim10^{3}\,M_\odot$ are included. Further extending the mass range to $10^{2}\,M_\odot$ produces negligible additional changes in $|F(f)|$. This behavior indicates that the effective mass range contributing to the WO signal in the LISA band is approximately $10^{4}$--$10^{7}\,M_\odot$ in our fiducial configuration.

The origin of this saturation can be understood in terms of characteristic time-delay scales. A subhalo of mass $m$ introduces a typical gravitational time delay of order $\Delta t \sim 4Gm/c^{3}$ (modulated by geometric factors near the macro image). For $m \lesssim 10^{3}\,M_\odot$, the associated time delays are too small to induce appreciable phase shifts across the LISA frequency band, and their contribution therefore remains effectively in the GO regime. Conversely, subhalos in the mass range $10^{4}$--$10^{7}\,M_\odot$ generate time-delay scales that overlap with $f\sim10^{-4}$--$10^{-1}\,\mathrm{Hz}$, allowing coherent WO distortions to develop. 

However, whether these perturbations translate into observable WO distortions depends sensitively on the macro configuration. The mere presence of subhalos is not sufficient to generate detectable frequency-dependent signatures. 
To isolate the role of the macro field, we repeated the calculation with the external quadratic term removed (i.e., $\psi_{\rm macro}=0$). In this case, the same subhalo realizations yield $|F(f)| \simeq 1$, with relative modulations suppressed to the $\lesssim 10^{-3}$ level (Appendix~E). The amplification factor is therefore nearly frequency-independent across the LISA band, indicating that the system remains effectively in the GO regime despite the presence of subhalos.

By contrast, in the strongly lensed configuration considered here, the macro image lies close to a critical curve, where the inverse Jacobian of the lens mapping is large. In this near-critical regime, small perturbations to the Fermat surface are geometrically amplified, stretching and redistributing the local time-delay structure. As a result, the projected subhalo population produces coherent WO distortions at the percent level over a broad frequency range.
These results demonstrate that macro critical amplification is not merely a quantitative enhancement, but the essential physical mechanism that elevates otherwise subdominant subhalo perturbations into observable WO signatures.

\section{Discussion}


\subsection*{Detectability of percent-level WO modulations}

Let $({\rm S/N})_{0}$ denote the intrinsic (unlensed) signal-to-noise ratio of the GW source. 
For a lensed event with amplification factor $|F|$, the observed signal-to-noise ratio scales as 
$|F|\,({\rm S/N})_{0}$. 
A fractional amplitude perturbation $\delta h/h$ is therefore detectable with an approximate significance 
$(\delta h/h)\, |F|\, ({\rm S/N})_0$, 
while phase perturbations become measurable once 
$\delta\phi \gtrsim [|F|\,({\rm S/N})_{0}]^{-1}$.

In our fiducial configuration, both the WO amplitude and phase modulations are typically of order $10^{-2}$ over part of the LISA band. 
For strongly magnified events with $|F|\sim 3$--$5$ and intrinsic $({\rm S/N})_{0}\gtrsim 100$, 
this implies an effective detection significance of several $\sigma$. 
Such intrinsic signal-to-noise ratios are routinely expected for massive black-hole binaries in the LISA band (e.g.,~\cite{LISA:2017pwj}), with favorable systems reaching $({\rm S/N})_{0}\sim 10^{3}$. 
For these loudest events, percent-level WO distortions would thus be detectable at high statistical significance.

\subsection*{Astrophysical relevance of small source offsets}

Our fiducial configuration adopts a dimensionless source offset
$y_{\rm src} = 0.1$.
In the SIS model,
sources with $0<y<1$ produce two images,
and a uniform source distribution in the source plane
implies a probability density, $p(y)=2y$ for $0\le y\le 1$,
so that
$P(y<y_0)=y_0^2$.
Naively, this gives
$P(y<0.1)=10^{-2}$,
i.e.\ only $\sim 1\%$ of sources lie within this offset.

However, detectable strongly lensed events are subject to
magnification bias.
In the SIS limit, the total magnification scales as
$\mu_{\rm tot}={2}/{y}$.
Since the GW strain amplitude scales as $\sqrt{\mu}$,
the horizon distance scales as $\sqrt{\mu}$,
and the accessible volume as $\mu^{3/2}$,
the effective detection-weighted distribution becomes
$
p_{\rm det}(y)\propto p(y)\,\mu^{3/2}
\propto y^{-1/2}
$.
Because $\mu$ formally diverges as $y\to 0$,
a finite maximum magnification $\mu_{\max}$
(or equivalently $y_{\min}=2/\mu_{\max}$)
must be imposed.
After normalization over $y\in[y_{\min},1]$,
the cumulative probability becomes
$
P_{\rm det}(y<y_0)
=
{(\sqrt{y_0}-\sqrt{y_{\min}})}
   /  ({1-\sqrt{y_{\min}}})
$.

For representative values $\mu_{\max}\sim 50$--$10^3$
(i.e.\ $y_{\min}\sim 0.04$--$0.002$),
one obtains
$P_{\rm det}(y<0.1)\sim 0.15\text{--}0.3$.
In the idealized SIS limit without imposing a maximum magnification
($\mu_{\max}\to\infty$), we obtain
$P_{\rm det}(y<0.1)=\sqrt{0.1}\simeq 0.32$.
Finite source size and deviations from the exact SIS profile
naturally reduce this value to the more conservative
$\sim 0.15$--$0.3$ range quoted above.
Thus, while $y<0.1$ represents only $\sim 1\%$
of sources in an unbiased population,
it can account for $\mathcal{O}(10\%)$
of detectable strongly lensed events
once magnification bias is included.

\section{Conclusions}

Strong gravitational lensing of GWs provides a qualitatively new probe of dark matter substructure through WO interference effects. 
We have performed full diffraction-integral calculations for strongly lensed GWs embedded in statistically generated cold dark matter subhalo populations down to $10^2\,M_\odot$, thereby quantifying the frequency-dependent amplification factor in the LISA band beyond the GO approximation.

We find that percent-level amplitude and phase modulations arise generically in galaxy-scale lenses when the macro image lies near a critical curve. 
Subhalos in the mass range $10^4$--$10^7\,M_\odot$ dominate the signal, as their characteristic time-delay scales naturally match the LISA band. 
In contrast, removing the macro critical field suppresses the frequency dependence to the $\lesssim 10^{-3}$ level, demonstrating that macro critical amplification is the essential mechanism that promotes otherwise subdominant perturbations into observable WO signatures.

For sufficiently magnified events with intrinsic signal-to-noise ratios $\gtrsim 10^2$--$10^3$, as expected for massive black-hole binaries detectable by LISA, these distortions should be measurable at high statistical significance. 
The configuration considered here is therefore not fine-tuned but representative of realistic strongly lensed GW events.

Remarkably, these signatures already arise within the standard cold dark matter paradigm without invoking compact objects or exotic substructure. 
Because the WO signal depends primarily on the compactness of perturbers and macro amplification near the image, any dark matter scenario producing more concentrated substructure---such as primordial black holes~\cite{Sasaki:2018dmp, Ando:2024ghr} or gravothermal core collapse in self-interacting dark matter~\cite{Tulin:2017ara,Ando:2024kpk}---would further enhance the effect.

Strongly lensed GWs in the LISA band thus provide a new and complementary window on dark matter structure at mass scales below those accessible to electromagnetic observations. 
Future work should extend the present analysis to other image configurations, particularly saddle images that are intrinsically more sensitive to perturbations, and to alternative dark matter scenarios with more compact substructure. 
Such extensions will further clarify the diagnostic power of WO distortions in strongly lensed GWs.

\begin{acknowledgments}
This work was supported by Grant-in-Aid for Scientific Research from the Ministry of Education, Culture, Sports, Science, and Technology (MEXT), Japan, grant numbers 20H05850, JP20H05861, and 24K07039. 
\end{acknowledgments}

\bibliographystyle{utphys}
\bibliography{refs}

\appendix

\section{Appendix~A. Subhalo populations from SASHIMI}

Subhalo populations are generated using the semi-analytic \textsc{SASHIMI} model~\cite{Hiroshima:2018kfv, Ando:2019xlm, Ando:2020yyk}. 
For a host halo with a specified mass and redshift, \textsc{SASHIMI} produces a catalog 
of subhalos characterized by their mass $m_{\rm sub}$, scale radius $r_s$, characteristic density 
$\rho_s$, and tidal radius $r_t$. Each catalog entry is associated with a weight $w$, 
which represents the ensemble-averaged number of subhalos with those properties expected 
within the host halo.

To construct macrolens realizations, we generate subhalos with masses 
$m_{\rm sub} > 10^9\,M_\odot$ using Monte Carlo sampling based on the weights $w$. 
The spatial distribution of subhalos within the host halo is assumed to follow 
a radial number density profile proportional to
\begin{equation}
n_{\rm sub}(r) \propto (r^2 + R_{s}^2)^{-3/2},
\end{equation}
where $R_{s}$ is the scale radius of the host halo. 
Subhalo positions are sampled from this distribution within the virial radius 
$R_{\mathrm{vir}}$ of the host halo, with angular positions drawn isotropically. 
The resulting three-dimensional distribution is then projected onto the lens 
plane to construct the macrolens realization used in the lensing calculations.

To model the WO perturbations produced by low-mass subhalos 
($m_{\rm sub} < 10^9\,M_\odot$), we focus on the region surrounding the macro image. 
For each entry in the \textsc{SASHIMI} catalog, the expected number of 
subhalos near the image is estimated from the fraction of the host halo 
volume intersecting a cylindrical region centered on the image.

Let $R_{\rm img}$ denote the projected distance of the macro image from 
the center of the host halo. We consider a cylinder of projected radius 
$R_{\rm near}$ around the image and integrate the assumed subhalo number 
density along the line of sight.  The line-of-sight 
integral of this profile within the virial radius $R_{\rm vir}$ gives
\begin{equation}
I_z = \int_{-z_{\rm max}}^{z_{\rm max}}
\frac{dz}{(z^2 + a^2)^{3/2}}
      = \frac{2 z_{\rm max}}{a^2 \sqrt{z_{\rm max}^2 + a^2}},
\end{equation}
where
$
z_{\rm max} = \sqrt{R_{\rm vir}^2 - R_{\rm img}^2}$ and
$a^2 = R_{\rm img}^2 + R_s^2$.
The effective volume of the cylindrical region around the image is then
\begin{equation}
\mathcal V_{\rm cyl} = \pi R_{\rm near}^2 I_z .
\end{equation}
This is normalized by the total effective volume of the host halo,
\begin{equation}
\mathcal V_{\rm halo} =
4\pi
\left[
\ln\!\left(
\frac{R_{\rm vir} + \sqrt{R_{\rm vir}^2 + R_s^2}}{R_s}
\right)
-
\frac{R_{\rm vir}}{\sqrt{R_{\rm vir}^2 + R_s^2}}
\right],
\end{equation}
which corresponds to integrating the same density profile over the 
entire halo volume.

For each catalog entry with weight $w$, the expected number of subhalos 
within the cylindrical region is therefore
\begin{equation}
w_{\rm cyl} = w \frac{\mathcal V_{\rm cyl}}{\mathcal V_{\rm halo}} .
\end{equation}
Subhalos are then generated using Monte Carlo sampling with this 
effective weight. Their projected positions are drawn uniformly within 
the disk of radius $R_{\rm near}$ centered on the macro image.

\section{Appendix~B. Derivation of the magnification--perturbation radius}
\label{app:Rmu}

In this Appendix, we derive the expression for the magnification--perturbation radius $R_{\mu}$ used in the main text.

In the GO limit, the magnification of an image is given by
\begin{equation}
\mu = \frac{1}{\det A},
\end{equation}
where $A$ is the Jacobian matrix of the lens mapping. Let $A_{\rm min}$ denote the Jacobian of the macrolens evaluated at the minimum image position. The corresponding magnification is $\mu_0 = 1/\det A_{\rm min}$.

A subhalo introduces a small perturbation to the lensing potential, leading to a correction $\delta A$ in the Jacobian:
\begin{equation}
A = A_{\rm min} + \delta A.
\end{equation}
For sufficiently small perturbations, the fractional change in magnification can be expanded to first order:
\begin{equation}
\frac{\delta \mu}{\mu_0}
\simeq
-\,\mathrm{Tr}\!\left(A_{\rm min}^{-1} \delta A \right).
\label{eq:dmu_linear}
\end{equation}

For a subhalo of mass $m_{\rm sub}$ located at projected separation $R$ from the image, the leading-order contribution to the convergence scales as
\begin{equation}
\delta \kappa(R) \sim \frac{m_{\rm sub}}{\pi R^2 \Sigma_{\rm cr}},
\end{equation}
where $\Sigma_{\rm cr}$ is the critical surface density. The induced perturbation in the Jacobian is therefore of order
\begin{equation}
\delta A \sim \delta \kappa \, I,
\end{equation}
up to order-unity geometric factors.

Substituting into Eq.~(\ref{eq:dmu_linear}), the fractional magnification perturbation becomes
\begin{equation}
\left| \frac{\delta \mu}{\mu_0} \right|
\sim
\|A_{\rm min}^{-1}\|
\frac{m_{\rm sub}}{\pi R^2 \Sigma_{\rm cr}},
\end{equation}
where $\|A_{\rm min}^{-1}\|$ characterizes the local macro-amplification and we define
\begin{equation}
g_{\rm img} \equiv \|A_{\rm min}^{-1}\|.
\end{equation}

We define the magnification--perturbation radius $R_{\mu}$ by requiring that the fractional perturbation exceed a threshold $\epsilon_\mu$:
\begin{equation}
\left| \frac{\delta \mu}{\mu_0} \right| \ge \epsilon_\mu.
\end{equation}
Solving for $R$ yields
\begin{equation}
R_{\mu}
=
\sqrt{
\frac{g_{\rm img}\, m_{\rm sub}}
{\pi \epsilon_\mu \Sigma_{\rm cr}}
}.
\end{equation}

This scale represents the maximum projected separation at which a subhalo of mass $m_{\rm sub}$ can induce a GO magnification perturbation of at least $\epsilon_\mu$ at the image location. In the main analysis, we adopt $\epsilon_\mu=10^{-2}$.

\section{Appendix~C. External-field decomposition in the WO integral}
\label{app:external}

We summarize how the external (macrolens) field is incorporated in the WO calculation through a local quadratic expansion around the macro minimum image.
We adopt the standard dimensionless formulation of wave-optics lensing~\cite{Takahashi:2003ix, Villarrubia-Rojo:2024xcj}, in which the amplification factor is
\begin{equation}
F(w)=\frac{w}{2\pi i}
\int d^2x\,
\exp\!\left[i w\,\phi(\bm{x},\bm{y})\right],
\label{eq:Fw_app}
\end{equation}
where $w$ is the dimensionless frequency and the dimensionless Fermat potential is
\begin{equation}
\phi(\bm{x},\bm{y})
=
\frac{1}{2}\left|\bm{x}-\bm{y}\right|^2
-\psi(\bm{x}).
\label{eq:fermat_app}
\end{equation}
Here $\bm{x}$ and $\bm{y}$ denote the dimensionless lens-plane and source coordinates defined in the main text, and $\psi$ is the dimensionless lensing potential.

Let $\bm{x}_{\rm min}$ denote the position of the macro minimum image, and define local coordinates
\begin{equation}
\bm{u}\equiv \bm{x}-\bm{x}_{\rm min}.
\end{equation}
A Taylor expansion of the Fermat potential gives
\begin{equation}
\phi(\bm{x},\bm{y})
\simeq
\phi_{\rm min}(\bm{y})
+
\left.\nabla\phi\right|_{\bm{x}_{\rm min}}\!\cdot \bm{u}
+
\frac{1}{2}\bm{u}^T
\left.\nabla\nabla\phi\right|_{\bm{x}_{\rm min}}
\bm{u} .
\label{eq:phi_expand_app}
\end{equation}
The linear term vanishes because $\bm{x}_{\rm min}$ is a stationary point of the Fermat potential in the GO limit,
\begin{equation}
\left.\nabla\phi\right|_{\bm{x}_{\rm min}}=0
\quad\Leftrightarrow\quad
\bm{y}
=
\bm{x}_{\rm min}-\left.\nabla\psi\right|_{\bm{x}_{\rm min}},
\end{equation}
i.e., the lens equation evaluated at the image.
The constant term $\phi_{\rm min}(\bm{y})$ contributes only an overall phase to $F(w)$ and will be dropped.

We decompose the total potential into a macrolens component and a local perturbative component,
\begin{equation}
\psi(\bm{x})=\psi_{\rm macro}(\bm{x})+\delta\psi(\bm{x}),
\end{equation}
where $\psi_{\rm macro}$ includes the host NFW halo, central SIS galaxy, and massive subhalos ($m_{\rm sub}>10^{9}\,M_\odot$),
while $\delta\psi$ is the contribution from nearby low-mass subhalos ($10^2$--$10^9\,M_\odot$) treated explicitly in the WO calculation.

Keeping the macrolens contribution to quadratic order around $\bm{x}_{\rm min}$ yields
\begin{equation}
\left.\nabla\nabla\phi\right|_{\bm{x}_{\rm min}}
=
I-\left.\nabla\nabla\psi_{\rm macro}\right|_{\bm{x}_{\rm min}}
\equiv A_{\rm min},
\label{eq:Amin_def}
\end{equation}
where $A_{\rm min}$ is the Jacobian matrix of the lens mapping evaluated at the minimum image.
It can be parameterized by an effective convergence $\kappa_{\rm macro}$ and shear components $(\gamma_{1,\rm macro},\gamma_{2,\rm macro})$ as
\begin{equation}
A_{\rm min}=
\begin{pmatrix}
1-\kappa_{\rm macro}-\gamma_{1,\rm macro} & -\gamma_{2,\rm macro}\\
-\gamma_{2,\rm macro} & 1-\kappa_{\rm macro}+\gamma_{1,\rm macro}
\end{pmatrix}.
\label{eq:Amin_kg_app}
\end{equation}

Inserting the quadratic expansion into Eq.~(\ref{eq:Fw_app}) gives, up to an irrelevant overall phase,
\begin{equation}
F(w)=\frac{w}{2\pi i}\int d^2u\,
\exp\!\left[
i w\left(
\frac{1}{2}\bm{u}^T A_{\rm min} \bm{u}
-\delta\psi(\bm{u})
\right)
\right].
\label{eq:Fw_local_app}
\end{equation}
In practice we treat the macrolens field entirely through $A_{\rm min}$, while the local subhalo potential is evaluated explicitly as a function of $\bm{u}$ (we suppress the dependence on $\bm{x}_{\rm min}$ for notational simplicity).

This construction ensures that: (i) the dominant macrolens field is incorporated exactly up to second order around the image, (ii) the stationary-point condition is preserved in the GO limit, and (iii) no double counting occurs between the macrolens field and the explicitly modeled local subhalo contribution.

\section{Appendix~D. Time-domain interpretation of the wave-optics amplification}

The amplification factor can equivalently be expressed in terms of the
distribution of the dimensionless Fermat potential.
Starting from
\begin{equation}
F(w)=\frac{w}{2\pi i}\int d^2x\, 
\exp\!\left[i w\,\phi(\bm{x})\right],
\end{equation}
one may insert the identity
$1=\int d\tau\,\delta\!\left(\tau-\phi(\bm{x})\right)$
to obtain
\begin{equation}
F(w)=\frac{w}{2\pi i}\int d\tau\, I(\tau)\, e^{i w \tau},
\label{eq:F_from_I_dimless}
\end{equation}
where
\begin{equation}
I(\tau)=\int d^2x\, 
\delta\!\left(\tau-\phi(\bm{x})\right)
\label{eq:I_dimless}
\end{equation}
denotes the distribution of the dimensionless Fermat potential.
Physically, $I(\tau)$ measures the area in the lens plane that contributes
with a given delay $\tau$.
Its normalization therefore reflects the local magnification of the image.

For a smooth macrolens in the GO limit,
the Fermat potential near a minimum image can be expanded as
\[
\phi(\bm{x}) \simeq 
\phi_{\rm min} + \frac{1}{2}\bm{u}^T A_{\rm min} \bm{u}.
\]
Because $A_{\rm min}$ is positive definite for a minimum,
the equal-delay contours are ellipses.
Evaluating Eq.~(\ref{eq:I_dimless}) explicitly for this quadratic form gives
\[
I_{0}^{\rm GO}(\tau)=\frac{2\pi}{\sqrt{\det A_{\rm min}}},
\]
i.e., a constant independent of $\tau$.
Since the macro magnification is $\mu_0=1/\det A_{\rm min}$,
the overall normalization of $I(\tau)$ is directly related to the magnification.
In this limit, $I(\tau)$ is smooth and featureless,
and Eq.~(\ref{eq:F_from_I_dimless}) yields a frequency-independent
amplification factor, as expected in GO.

\begin{figure}[t]
\centering
\includegraphics[width=\columnwidth]{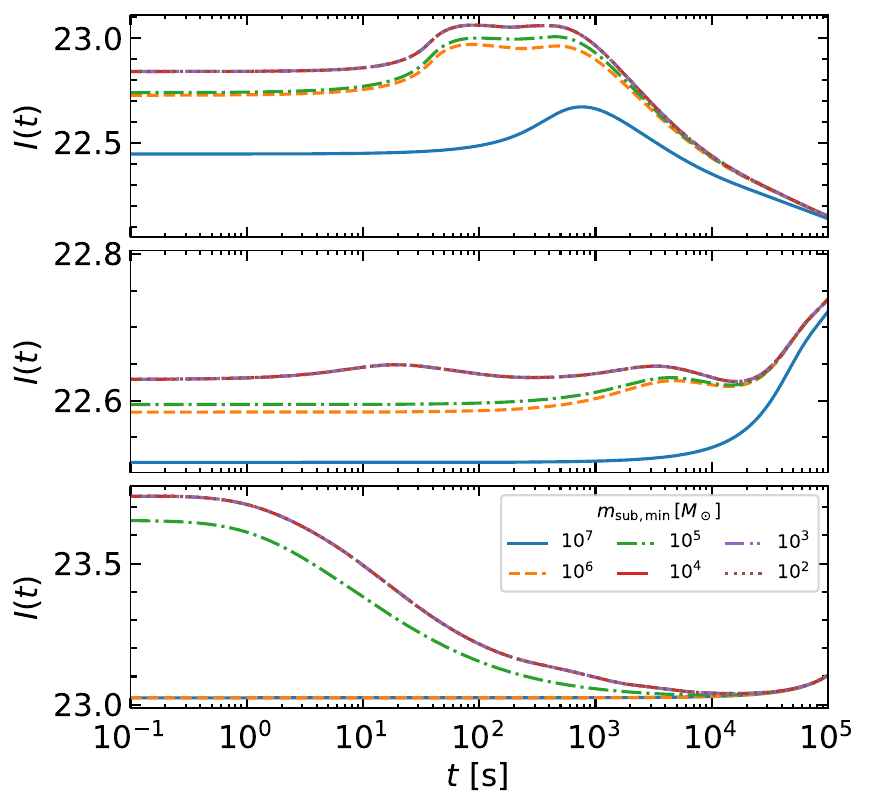}
\caption{
Distribution of the dimensionless Fermat potential $I(\tau)$
for three representative realizations,
shown for different minimum subhalo masses
$m_{\rm sub,min}=10^{4}$--$10^{7}\,M_\odot$ (see legend).
In the pure geometric-optics limit,
$I(\tau)$ would be constant (for a quadratic minimum)
with normalization proportional to the macro magnification.
Including progressively lower-mass subhalos introduces
localized distortions in the shape of $I(\tau)$,
while leaving its overall normalization largely unchanged.
These distortions are responsible for the frequency-dependent
structure of the amplification factor $F(w)$.
}
\label{fig:It_msubmin}
\end{figure}

The inclusion of subhalos perturbs the local structure of the Fermat surface,
introducing small distortions in $\phi(\bm{x})$ near the minimum image.
These distortions modify the detailed shape of $I(\tau)$,
as illustrated in Fig.~\ref{fig:It_msubmin}.
While the overall normalization of $I(\tau)$ remains primarily determined
by the macrolens (the values $I(\tau)\simeq 22$--$24$ in the figure
correspond to the integrated area associated with the macro minimum
in our normalization),
the profile develops increasingly pronounced structure
as progressively lower-mass subhalos are included.
Because $F(w)$ is the Fourier transform of $I(\tau)$,
such distortions in the time-domain distribution translate directly
into frequency-dependent modulations in the amplification factor.

This time-domain picture makes explicit the nonlinear nature of the
WO signal:
subhalos collectively reshape the distribution of the Fermat potential,
whose Fourier response determines the observable modulation.

\section{Appendix E. WO calculation without the external macro field}

To isolate the physical role of the external macro field in generating observable WO signatures, we repeat the calculation with the quadratic external term switched off in the diffraction integral. In this setup, only the local subhalo potentials are included in the lens model, while the host halo, central galaxy, and massive subhalos treated in the GO limit are removed from the phase. This comparison allows us to determine whether the frequency-dependent structures reported in the main text arise intrinsically from the subhalo population itself, or from its interplay with the macro criticality of the strongly lensed configuration.

\begin{figure}[t]
\centering
\includegraphics[width=\columnwidth]{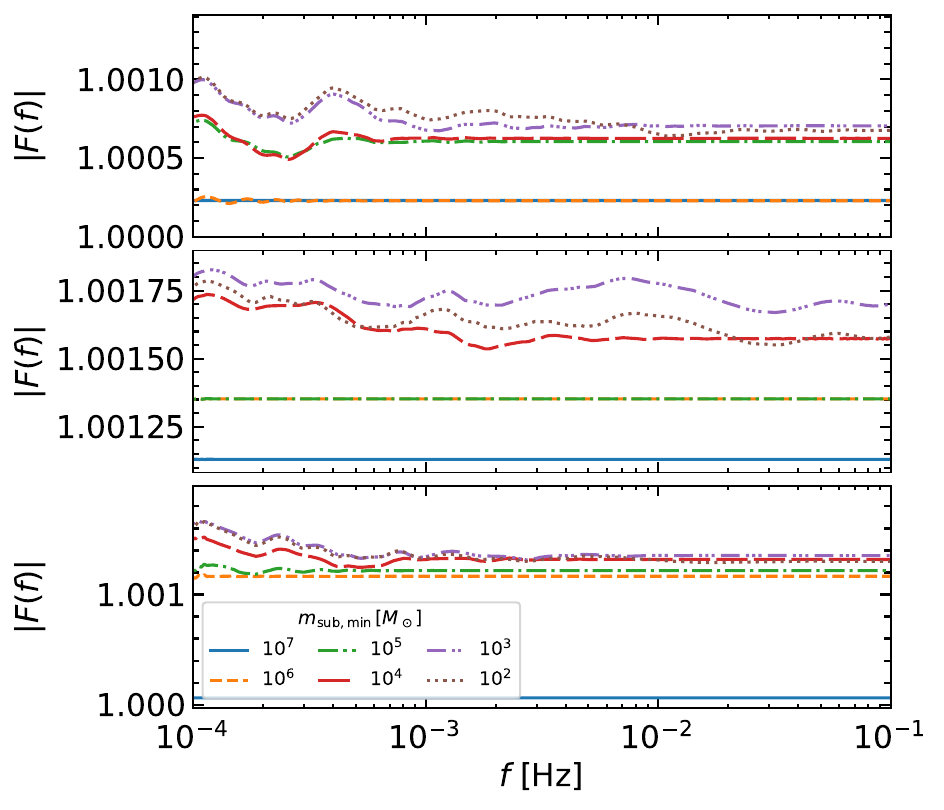}
\caption{
Frequency-domain amplification factor $|F(f)|$ computed without the external macro field.
The same three realizations and mass thresholds as in Fig.~\ref{fig:Fw_msubmin} are shown.
Across the LISA band, the amplification factor is nearly frequency-independent,
with only weak broadband curvature and no prominent oscillatory structure.
}
\label{fig:Fw_msubmin_no_ext}
\end{figure}

The corresponding frequency-domain amplification factor is shown in Fig.~\ref{fig:Fw_msubmin_no_ext}. 
Across the LISA band, $|F(f)|$ remains nearly frequency-independent in all realizations, with relative modulations suppressed to the $\lesssim 10^{-3}$ level. 
Variations induced by changing $m_{\rm sub,min}$ are small. 
Even subhalos in the mass range $10^{4}$--$10^{7}\,M_{\odot}$, which dominate the WO signal in the near-critical configuration discussed in the main text, fail to generate significant frequency-dependent structure once the external macro field is removed.

This behavior follows directly from Eq.~(\ref{eq:Ff}). 
With the external macro potential switched off, the local Jacobian reduces to $A_{\rm min}=I$, and the diffraction integral is governed primarily by the quadratic term $\bm{u}^T\bm{u}/2$. 
The subhalo potential $\delta\psi(\bm{u})$ then acts only as a small perturbation to an otherwise isotropic quadratic phase, so the Fermat surface remains well approximated by a local quadratic form. 
The integral therefore reduces effectively to its stationary-phase (GO) limit, yielding an amplification factor that is nearly constant across the LISA band.

These results confirm that the percent-level WO distortions reported in the main text arise from the interplay between local subhalo perturbations and macro critical amplification. 
Near a critical curve, the large inverse macro-Jacobian enhances small deviations from quadratic structure in the Fermat surface, redistributing the local time-delay pattern and promoting otherwise subdominant perturbations into observable WO signatures. 
In the absence of this macro criticality, the same subhalo realization remains effectively in the GO regime.

\section{Appendix~F. Additional realizations and source configurations}

\begin{figure}[t]
\centering
\includegraphics[width=\columnwidth]{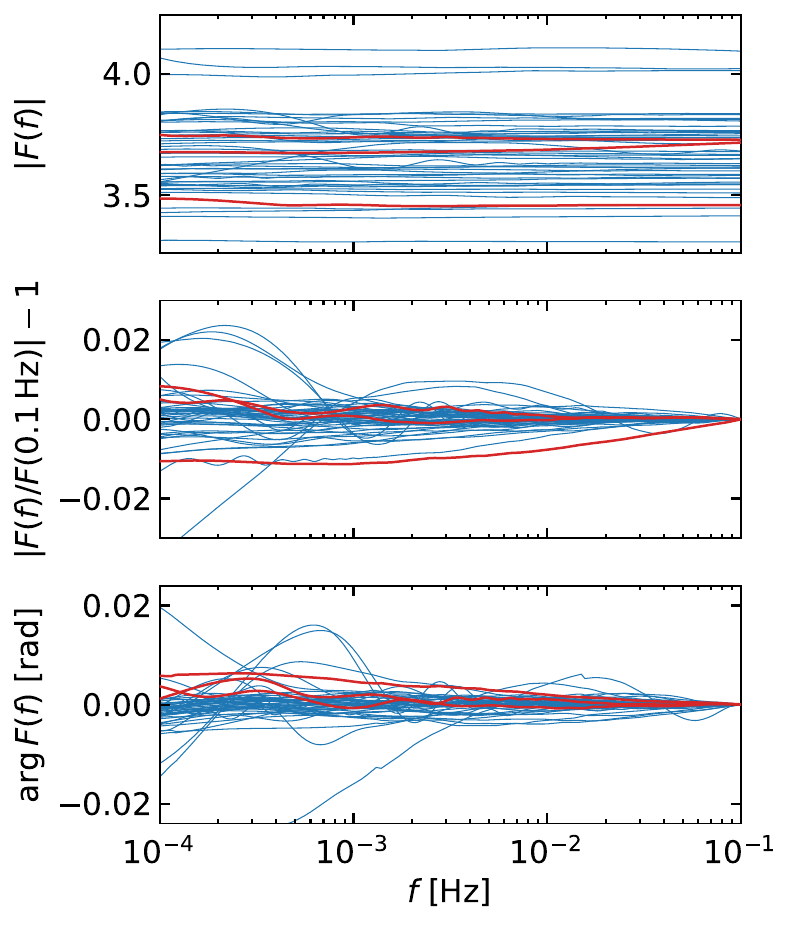}
\caption{
Amplification factors for 50 individual realizations of the subhalo population.
The top panel shows the absolute amplification $|F(f)|$, the middle panel
displays the fractional modulation $|F(f)/F(0.1\,\mathrm{Hz})|-1$, and the
bottom panel presents the phase shift $\arg F(f)$. Blue curves correspond
to different Monte Carlo realizations of the subhalo distribution, while
three representative realizations (the same as in the main text) are highlighted in red. Although the
detailed interference pattern varies between realizations, the overall
modulation amplitude and characteristic frequency scale remain similar.
}
\label{fig:Fw_individual}
\end{figure}

Figure~\ref{fig:Fw_individual} illustrates 50 individual realizations of the
subhalo population around the macro minimum image. Each curve corresponds to a
different Monte Carlo realization generated using the procedure described in
Appendix~A. The detailed oscillatory structure of the amplification factor
varies between realizations because it depends sensitively on the positions
and masses of nearby subhalos. Nevertheless, the overall scale of the
frequency-dependent modulation remains broadly similar across realizations.
These examples demonstrate the stochastic nature of the WO signatures
induced by subhalos while confirming that percent-level distortions in both
amplitude and phase arise generically in strongly lensed configurations.

\begin{figure}[t]
\centering
\includegraphics[width=\columnwidth]{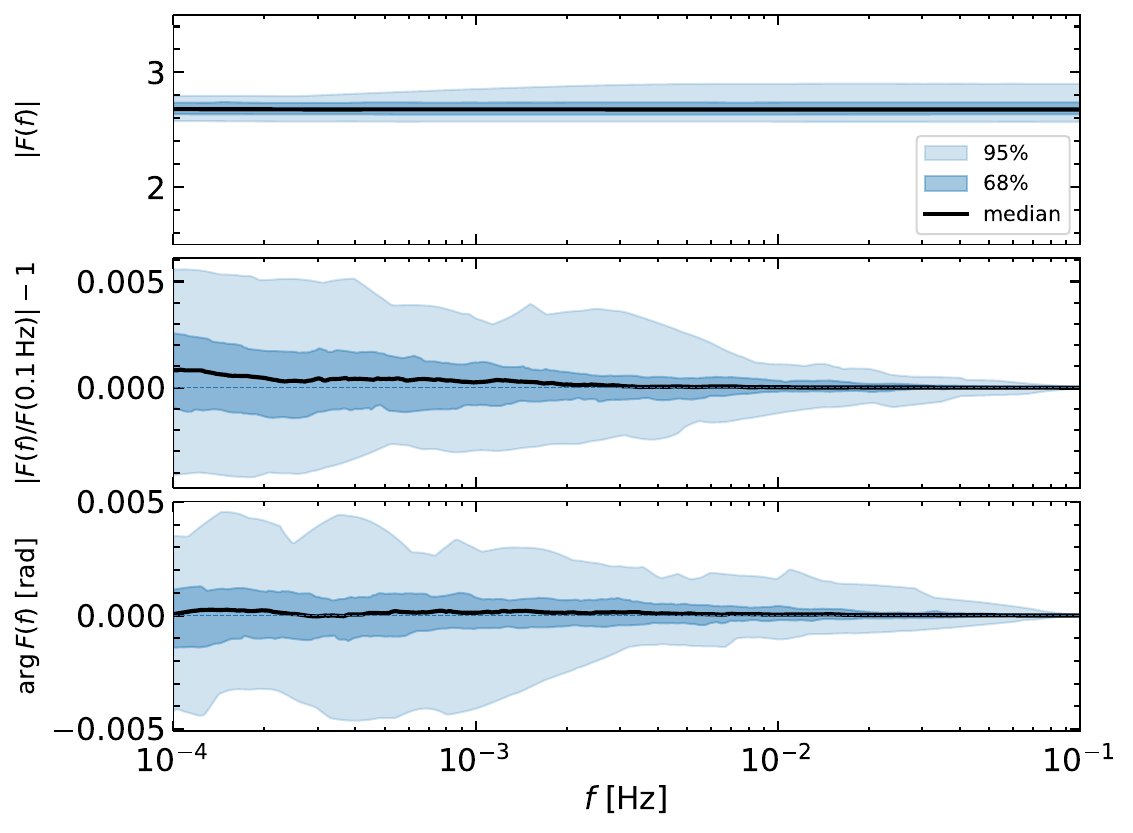}
\caption{
The same as Fig.~\ref{fig:Fw_percentiles}, but for $y_{\rm src}=0.2$.
}
\label{fig:Fw_percentiles_y02}
\end{figure}

Figure~\ref{fig:Fw_percentiles_y02} shows the same percentile statistics
as Fig.~\ref{fig:Fw_percentiles}, but for a larger source offset
$y_{\rm src}=0.2$. As expected, the typical modulation amplitude becomes
smaller than in the $y_{\rm src}=0.1$ case because the macro image lies
farther from the critical curve and therefore has a lower magnification.

Although larger values of $y_{\rm src}$ correspond to a larger source-plane
area, detectable strongly lensed events are subject to magnification bias.
Using the detection-weighted distribution derived in the main text,
$p_{\rm det}(y)\propto y^{-1/2}$, the cumulative probability scales
approximately as $P_{\rm det}(y<y_0)\propto \sqrt{y_0}$ (for
$y_0 \gg y_{\min}$). This implies
\[
\frac{P_{\rm det}(y<0.2)}{P_{\rm det}(y<0.1)}
\simeq
\frac{\sqrt{0.2}}{\sqrt{0.1}}
\simeq
1.4 .
\]
Thus, extending the source offset from $y<0.1$ to $y<0.2$ increases the
expected detection rate only by a factor of $\sim1.4$, while the
WO modulation becomes significantly smaller. Consequently,
events with $y_{\rm src}\lesssim0.1$ remain particularly important for
detecting WO signatures from dark matter subhalos.

\end{document}